\documentclass[
    aps,                 
    prd,                 
    reprint,             
    superscriptaddress,  
    nofootinbib,         
    eprint,              
]{revtex4-2}

\usepackage{graphicx}   
\usepackage{subfigure}  
\usepackage{dcolumn}    
\usepackage{bm}         
\usepackage{multirow}   
\usepackage{float}      

\usepackage{amsmath}
\usepackage{amssymb}
\usepackage{amsthm}

\usepackage{hyperref}
\hypersetup{
    colorlinks=true,    
	linkcolor=blue,     
	citecolor=green,    
	filecolor=magenta,  
	urlcolor=cyan,      
}

\usepackage{cleveref}
\makeatletter
\AddToHook{cmd/appendix/before}{\def\cref@section@alias{appendix}}  
\makeatother

\usepackage{glossaries-extra}  
\glsdisablehyper               
\newglossaryentry{BH}{
    name=BH,
    description=Black Hole,
    first=Black Hole (BH),
    plural=BHs,
    descriptionplural=Black Holes,
    firstplural=Black Holes (BHs)
}

\newglossaryentry{BS}{
    name=BS,
    description=Boson Star,
    first=Boson Star (BS),
    plural=BSs,
    descriptionplural=Boson Stars,
    firstplural=Boson Stars (BSs)
}

\newglossaryentry{EM}{
    name=EM,
    description=electromagnetic,
    first=electromagnetic (EM)
}

\newglossaryentry{GR}{
    name=GR,
    description=General Relativity,
    first=General Relativity (GR)
}

\newglossaryentry{EoM}{
    name=EoM,
    description=equation of motion,
    first=equation of motion (EoM),
    plural=EoM,
    descriptionplural=equations of motion,
    firstplural=equations of motion (EoM)
}

\newglossaryentry{GW}{
    name=GW,
    description=gravitational wave,
    first=gravitational wave (GW),
    plural=GWs,
    descriptionplural=gravitational waves,
    firstplural=gravitational waves (GWs)
}

\newglossaryentry{LISA}{
    name=LISA,
    description=Laser Interferometer Space Antenna,
    first=Laser Interferometer Space Antenna (LISA)
}

\newglossaryentry{EMS}{
    name=EMS,
    description=Einstein-Maxwell-Scalar,
    first=Einstein-Maxwell-Scalar (EMS)
}

\usepackage{lipsum}

\begin{document}

\title{Spontaneous spherical symmetry breaking of black holes with resonant hair}


\author{José Ferreira}
\email{jpmferreira@ua.pt}
\affiliation{Departamento de Matemática da Universidade de Aveiro and Centre for Research and Development in Mathematics and Applications (CIDMA), Campus de Santiago, 3810-193 Aveiro, Portugal}

\author{Carlos A. R. Herdeiro}
\email{herdeiro@ua.pt}
\affiliation{Departamento de Matemática da Universidade de Aveiro and Centre for Research and Development in Mathematics and Applications (CIDMA), Campus de Santiago, 3810-193 Aveiro, Portugal}

\author{Eugen Radu}
\email{eugen.radu@ua.pt}
\affiliation{Departamento de Matemática da Universidade de Aveiro and Centre for Research and Development in Mathematics and Applications (CIDMA), Campus de Santiago, 3810-193 Aveiro, Portugal}

\author{Miguel Zilhão}
\email{mzilhao@ua.pt}
\affiliation{Departamento de Física da Universidade de Aveiro and Centre for Research and Development in Mathematics and Applications (CIDMA), Campus de Santiago, 3810-193 Aveiro, Portugal}

\date{April 2026}

\begin{abstract}
    Black holes with resonant hair are static, spherical, electrically charged solutions of the Einstein-Maxwell-(gauged-)scalar system. Scalar self-interactions are mandatory for their existence. Initial dynamical studies restricted to spherical symmetry suggested stability; more recently, fully non-spherical dynamical studies revealed instabilities, at least for a particular class of self-interactions. Here, we provide a more detailed study of this instability together with a different decay channel, depending on the chosen solutions. Moreover, considering a second model, we provide evidence that the instabilities may be generic for different classes of self-interactions. We conclude these solutions are dynamically unstable and split into a bosonic lump and a bald black hole (via fission) or implode to the latter (via absorption). In both cases, the non-spherical dynamics seems to be key. 
\end{abstract}

\maketitle


\section{Introduction}
\label{sec:introduction}

Since the initial detections of \glspl{GW} in 2015 \cite{GW190521}, four observation catalogs have been released, totaling over $200$ \gls{GW} events \cite{GWTC-1,GWTC-2,GWTC-3,LIGOScientific:2025hdt}. In the coming years, the detection rate is anticipated to increase substantially, with both improvements to current detectors but also due to forthcoming next generation observatories such as \gls{LISA}~\cite{LISA} and the Einstein Telescope. These advancements in \gls{GW} astronomy will enable more precise tests of \gls{GR} and the nature of compact objects.

Simultaneously, advances in Numerical Relativity have allowed for the creation of larger and increasingly more accurate catalogs of numerically generated waveforms \cite{RIT-Catalog,SXS-Catalog,Georgia-Catalog,NRAR-Catalog}. Notably, these catalogs are generated exclusively from the merger of Kerr \glspl{BH}. Although these have been in good agreement with observations so far, to probe possible deviations from the Kerr paradigm, it becomes imperative to include waveforms from alternative models to future catalogs. A pioneering effort in this direction has been reported and used in~\cite{Bustillo2021,Sanchis-Gual:2022mkk}.

Several uniqueness theorems constrain the properties of \glspl{BH}. However, these theorems can be evaded in multiple ways and create what are known as hairy \gls{BH} solutions -- see \emph{e.g.}~\cite{Herdeiro2015a}. Nevertheless, for any such object to serve as a viable alternative to the Kerr \gls{BH}, stability over astrophysically relevant timescales is required.

A notable example is \glspl{BH} with resonant scalar hair~\cite{Herdeiro2020}. These are static, spherically symmetric solutions to the \gls{EMS} model with a self-interacting, gauged, and minimally coupled scalar field. Recent nonlinear evolutions indicate that these solutions can be dynamically stable in spherical symmetry~\cite{Zhang2023,Zhao2025}, and can form via non-linear interaction with a charged scalar field~\cite{Zhang2023}. 

This work investigates the nonlinear evolution of \glspl{BH} endowed with resonant scalar hair beyond spherical symmetry, in 3+1 numerical relativity, evolving different configurations and exploring the possible end states, following up on the work~\cite{Nicoules2025}. The main conclusion is that these solutions are unstable, when non-spherical dynamics is taken into account. Moreover, we unveil a different decay channel from the one described in~\cite{Nicoules2025}, show that the results are similar for different types of scalar potentials and conduct an analysis of the decay timescales for the unstable configurations. 

The structure of this paper is as follows. \Cref{sec:model} introduces the \gls{EMS} model, details the equations of motion, and describes the 3+1 decomposition. In \cref{sec:initial-data} we outline the procedure for constructing initial data and the properties of representative configurations across the parameter space. \Cref{sec:results} presents the numerical setup and an analysis of the results. Final remarks and perspectives are provided in \cref{sec:final-remarks}. Two appendices provide further technical details.

\section{Model}
\label{sec:model}

\subsection{Action and Field Equations}
\label{subsec:model-action}

We consider an EMS model described
by the Lagrangian density
\begin{equation}
    \label{eq:action}
    \mathcal{L} =
        \frac{R}{16 \pi} -
        \frac{1}{4} F_{\alpha \beta} F^{\alpha \beta} -
        (\widetilde{\nabla}_\alpha \phi)^* \widetilde{\nabla}^\alpha \phi - V(\left| \phi \right|)
    \,,
\end{equation}
where $R$ denotes the Ricci scalar, $F_{\mu \nu} \equiv \nabla_\mu A_\nu - \nabla_\nu A_\mu$ is the Maxwell tensor, $A_\mu$ is the electromagnetic four-potential, $\phi$ is the complex scalar field, and $V(\left| \phi \right|)$ its potential. The (minimal) coupling between the scalar and the electromagnetic field is introduced via the gauge covariant derivative
\begin{equation}
    \label{eq:coupled-derivative}
    \widetilde{\nabla}_\mu \phi \equiv \nabla_\mu \phi + i q A_\mu \phi \,,
\end{equation}
where $q$ is the coupling constant.

Variation of the action with respect to the metric yields the Einstein field equations
\begin{equation}
    \label{eq:EFE}
    R_{\mu \nu} - \frac{1}{2} g_{\mu \nu} R =
    8 \pi \left( T_{\mu \nu}^\text{EM} + T_{\mu \nu}^\phi \right) \,,
\end{equation}
where $T_{\mu \nu}^\text{EM}$ and $T_{\mu \nu}^\phi$ are the energy-momentum tensors for the electromagnetic and scalar field, respectively, given by
\begin{subequations}
\begin{align}
    \label{eq:Tmunu-EM}
    T_{\mu \nu}^\text{EM} &=
    F_\mu{}^{\alpha} F_{\nu \alpha} - \frac{1}{4} g_{\mu \nu} F_{\alpha \beta} F^{\alpha \beta} \,,
    \\
    \label{eq:Tmunu-Psi}
    T_{\mu \nu}^\phi &=
    2\widetilde{\nabla}_{(\mu} \phi^* \widetilde{\nabla}_{\nu)} \phi +
    g_{\mu \nu} \left[ \widetilde{\nabla}^\alpha \phi^* \widetilde{\nabla}_\alpha \phi + V(\left| \phi \right|) \right] \,.
\end{align}
\end{subequations}

Variation with respect to the electromagnetic potential leads to the Maxwell equations
\begin{equation}
    \label{eq:EoM-EM}
    \nabla_\alpha F^{\mu \alpha} = j^\mu \,,
\end{equation}
where the electromagnetic current sourced by the scalar field is
\begin{equation}
    \label{eq:EM-current}
    j^\mu = i q \left[ (\widetilde{\nabla}^\mu \phi)^* \phi - \phi^* (\widetilde{\nabla}^\mu \phi) \right] \,.
\end{equation}

Variation with respect to the scalar field yields
\begin{equation}
    \label{eq:EoM-scalar}
    \widetilde{\nabla}_\alpha \widetilde{\nabla}^\alpha \phi =
    \frac{dV}{d\left| \phi \right|^2} \phi \, .
\end{equation}
We consider two different choices for $V(|\phi|)$: (i) a $Q$-ball type potential
\begin{equation}
    \label{eq:sextic-potential}
    V(|\phi|) = \mu^2 |\phi|^2 (1 - 2 \lambda |\phi|^2)^2 \,,
\end{equation}
where $\mu$ is the mass term and $\lambda$ a self-interaction parameter; (ii) the axionic potential
\begin{equation}
    \label{eq:axionic-potential}
    V(|\phi|) = \frac{2 \mu^2 f^2}{B} \left( 1 - \sqrt{1 - 4 B \sin \left( \frac{|\phi|}{2f} \right)} \right) \,,
\end{equation}
where $f$ and $B$ are free parameters, with $4|B|\leqslant 1$.

\subsection{Cauchy problem formulation}
\label{subsec:model-cauchy}

To solve the previous equations numerically, we rewrite them using the $3+1$ decomposition \cite{Baumgarte2010,Alcubierre2009}.
The spacetime line element is written as%
\footnote{Greek indices denote four-dimensional quantities, while Latin indices refer to three-dimensional spatial components.}
\begin{equation}
    \label{eq:3+1-line-element}
    ds^2 = -\alpha^2 dt^2 + \gamma_{ij}(dx^i + \beta^i dt)(dx^j + \beta^j dt) \,,
\end{equation}
where $\alpha$ is the lapse, $\beta^i$ is the shift vector and $\gamma_{\mu \nu}$ is the induced metric on the spatial hypersurface, defined as
\begin{equation}
    \label{eq:3-metric}
    \gamma_{\mu \nu} \equiv g_{\mu \nu} + n_\mu n_\nu \,,
\end{equation}
with $n_\mu = (-\alpha,0,0,0)$ the future-directed unit normal vector.

The electromagnetic four-potential is decomposed into spatial and normal components,
\begin{equation}
    A_\mu = \mathcal{A}_\mu + A_\phi n_\mu \,,
\end{equation}
where $\mathcal{A}_\alpha \equiv \gamma_\alpha{}^\beta A_\beta$ and $A_\phi \equiv -n^\alpha A_\alpha$. The Maxwell tensor becomes
\begin{equation}
    \label{eq:Fmunu-decomposed}
    F_{\mu \nu} = D_\mu \mathcal{A}_\nu - D_\nu \mathcal{A}_\mu + n_\mu E_\nu - n_\nu E_\mu \,,
\end{equation}
where $E^\mu \equiv \gamma^{\mu \alpha} n^\beta F_{\alpha \beta}$ is the electric field, and $D_i$ is the covariant derivative compatible with $\gamma_{ij}$, defined generically as
\begin{align}
    \label{eq:3D-derivative}
    \begin{split}
        & D_i T^{\mu_1 \ldots \mu_n}{}_{\nu_1 \ldots \nu_n} \equiv \\
        & \gamma^{\mu_1}{}_{\alpha_1} \ldots \gamma^{\mu_n}{}_{\alpha_n}
        \gamma^{\beta_1}{}_{\nu_1}  \ldots \gamma^{\beta_n}{}_{\nu_n}
        \gamma^\rho{}_i
        \nabla_\rho T^{\alpha_1 \ldots \alpha_n}{}_{\beta_1 \ldots \beta_n} \,.
    \end{split}
\end{align}

Contracting the first index of \cref{eq:Fmunu-decomposed} with $n_\mu$ and projecting the second index with $\gamma^\nu{}_\beta$ yields the evolution equation for $\mathcal{A}^i$,
\begin{equation}
    \label{eq:EoM-Ai}
    \partial_t \mathcal{A}^i =
    - \alpha E^i
    - D^i(A_\phi \alpha )
    + \beta^j D_j \mathcal{A}^i - \mathcal{A}^j D_j \beta^i
    \,.
\end{equation}
Substituting \cref{eq:Fmunu-decomposed} into \cref{eq:EoM-EM}, the normal projection gives the constraint
\begin{equation}
    \label{eq:EM-constraint}
    M \equiv D_i E^i - 4 \pi \rho_e = 0 \,,
\end{equation}
where $\rho_e \equiv - n_\mu j^\mu$ is the normal component of the four-current. The spatial projection yields the evolution equation
\begin{equation}
\begin{split}
    \label{eq:EoM-Ei}
   \partial_t E^i &=
        \alpha K E^i
        + \beta^j D_j E^i - E^j D_j \beta^i
        \\ &
        + D_j (\alpha (D^i \mathcal{A}^j - D^j \mathcal{A}^i) )
        - \alpha z^i
    \,,
\end{split}
\end{equation}
where $z^i \equiv \gamma^i{}_k j^k$ is the spatial projection of $j^\mu$, and $K \equiv \gamma^{ij} K_{ij}$ is the trace of the extrinsic curvature. The $3+1$ decomposition of the current yields
\begin{subequations}
\begin{align}
    \rho_e &= - 2 i q (\phi K_\phi^* - \phi^* K_\phi)
              - 2 q^2 A_\phi \phi \phi^* \,,
    \\
    z^i &= - 2 i q \left( \phi \partial^i \phi^* - \phi^* \partial^i \phi \right)
           - 2 q^2 \mathcal{A}^i \phi \phi^*
    \,,
\end{align}
\end{subequations}
where the conjugate momentum of the scalar field $K_\phi$ is defined as
\begin{equation}
    \label{eq:K_phi}
    K_\phi \equiv - \frac{1}{2} n^\alpha \nabla_\alpha \phi \,.
\end{equation}

The Lorenz gauge condition, $\nabla_\mu A^\mu = 0$, is imposed to fix the electromagnetic gauge freedom, leading to the evolution equation for $A_\phi$,
\begin{equation}
    \label{eq:3+1-EoM-gauge}
    \partial_t A_\phi =
        \beta^i \partial_i A_\phi
        + \alpha K A_\phi
        - \alpha D_j \mathcal{A}^j - \mathcal{A}^j \partial_j \alpha
    \,.
\end{equation}

To control violations of the constraint \cref{eq:EM-constraint}, the equations of motion are modified following the approach of \cite{Zilhao2015,Hilditch2013}, introducing an auxiliary variable $Z$ and a constraint damping parameter $\kappa$. The modified equations read
\begin{subequations}
\begin{align}
    \begin{split}
        \label{eq:3+1-EoM-constraint-E}
        \partial_t E^i &=
            \alpha K E^i
            + \beta^j D_j E^i - E^j D_j \beta^i \\
            &+ D_j (\alpha (D^i \mathcal{A}^j - D^j \mathcal{A}^i) )
            + \alpha D^i Z
        - 4 \pi \alpha z^i
        \,,
    \end{split}
    \\
    \begin{split}
        \label{eq:3+1-EoM-constraint-gauge}
        \partial_t A_\phi &=
            \beta^i \partial_i A_\phi
            + \alpha K A_\phi \\
            &- \alpha D_j \mathcal{A}^j - \mathcal{A}^j \partial_j \alpha
            - \alpha Z
        \,.
    \end{split}
\end{align}
\end{subequations}
with the auxiliary variable $Z$ evolving as
\begin{equation}
    \label{eq:3+1-EoM-constraint-Z}
    \partial_t Z =
        \alpha M
        - \alpha \kappa Z
        + \beta^j \partial_j Z
    \,.
\end{equation}

For the scalar field, a first-order reduction is performed,
\begin{equation}
    \nabla_\mu \phi = D_\mu \phi + 2 K_\phi n_\mu \,,
\end{equation}
where $K_\phi$ is taken as an independent variable. Rewriting \cref{eq:K_phi} we obtain the time derivative for the scalar field
\begin{equation}
    \label{eq:EoM-phi}
    \partial_t \phi = \beta^j \partial_j \phi - 2 \alpha K_\phi \,.
\end{equation}
Substituting \cref{eq:K_phi} into \cref{eq:EoM-scalar} gives the evolution equation for $K_\phi$,
\begin{equation}
\begin{split}
    \label{eq:EoM-Kphi}
    \partial_t K_\phi &=
    \alpha K K_\phi +
    \beta^j \partial_j K_\phi
    - \frac{1}{2} D^i(\alpha \partial_i \phi)
    + \frac{1}{2} \alpha \frac{dV}{d\left| \phi \right|^2} \phi
    \\ &
    - i \alpha q \left( \mathcal{A}^i \partial_i \phi - 2 K_\phi A_\phi \right)
    + \frac{\alpha q^2}{2} \left( \mathcal{A}^i \mathcal{A}_i - A_\phi^2 \right) \phi
    \,,
\end{split}
\end{equation}

\subsection{Physical Quantities}
\label{subsec:physical-quantities}

The minimal coupling between the scalar and electromagnetic fields means the scalar field is electrically charged. A quantum of scalar field has charge $q$. The total charge of a BH-scalar field system can then be decomposed as \cite{Herdeiro2020}
\begin{equation}
    \label{eq:decomposition-Q}
    Q = Q_h + Q_\phi \,,
\end{equation}
where $Q_\phi$ is the total charge associated with the scalar field and $Q_h$ is the charge associated with the horizon region. In the $3+1$ formalism, the scalar field charge is computed as
\begin{equation}
    \label{eq:Q_phi}
    Q_\phi = \frac{1}{4 \pi} \int_{r_h}^\infty \sqrt{\gamma} \rho_e  \, d^3x \,,
\end{equation}
where $\gamma$ is the determinant of the spatial metric and $\rho_e$ is the normal projection of the electromagnetic current defined above. $r_h$ is the radial coordinate of the BH horizon.

As for the BH charge, $Q_h$, it is defined via a surface integral at the horizon,
\begin{equation}
    Q_h = - \frac{1}{4 \pi} \oint_{\mathcal{H}} \sqrt{\gamma^{(2)}} F^{\mu \nu} n_\mu \sigma_\nu \, d^2x \,,
\end{equation}
where $\gamma^{(2)}$ is the determinant of the induced two-metric on the horizon $\mathcal{H}$, $n_\mu$ is the future-directed unit normal to the hypersurface, and $\sigma_\nu$ is the outward-pointing normal to the horizon surface. To avoid computing surface integrals, we determine the value of $Q_h$ by computing the total charge $Q$ (as explained next) and using \cref{eq:decomposition-Q}.

To evaluate the total charge and the outgoing electromagnetic radiation, we make use of the Newman–Penrose scalars defined as \cite{Newman1962}
\begin{subequations}
\begin{align}
    \label{eq:Phi1}
    \Phi_1 & \equiv \frac{1}{2} F_{\mu \nu} \left(l^{\mu} k^{\nu} + \bar m^{\mu} m^{\nu} \right) \,,
    \\
    \label{eq:Phi2}
    \Phi_2 & \equiv F_{\mu \nu} \bar m^{\mu} k^{\nu} \,,
\end{align}
\end{subequations}
where $l$, $k$ and $m$ are constructed from an orthonormal triad $u$, $v$, $w$ orthogonal to $n^{\mu}$ according to
\begin{subequations}
\begin{align}
    l^{\alpha} &= \frac{1}{\sqrt{2}} \left( n^{\alpha} +   u^{\alpha} \right) \,, \\
    k^{\alpha} &= \frac{1}{\sqrt{2}} \left( n^{\alpha} -   u^{\alpha} \right) \,, \\
    m^{\alpha} &= \frac{1}{\sqrt{2}} \left( v^{\alpha} + i w^{\alpha} \right) \,.
\end{align}
\end{subequations}

At sufficiently large coordinate radius $R_\text{ex}$ the scalars are decomposed in spin-weighted spherical harmonics
\begin{subequations}
\begin{align}
    \label{eq:Phi1_decomposed}
    \Phi_1 &= \sum_{l,m} \phi_{1}^{lm}(t) Y_{lm}^{0}  \,, \\
    \Phi_2 &= \sum_{l,m} \phi_{2}^{lm}(t) Y_{lm}^{-1} \,,
\end{align}
\end{subequations}
which, for outgoing waves in the far-field regime, are related to the components of the electric and magnetic fields as \cite{Zilhao2012}
\begin{subequations}
\begin{align}
    \Phi_1 & \approx \frac{1}{2} \left(E_{\hat{r}} + i B_{\hat{r}} \right) \,, \\
    \Phi_2 & \approx E_{\hat{\theta}} - i E_{\hat{\varphi}} \,.
\end{align}
\end{subequations}
For a non rotating, point-like, charged particle in flat spacetime, the only non-zero component of the electric field is 
\begin{equation}
    E_{\hat{r}} =  \frac{ Q}{r^2} \,,
\end{equation}
which when inserted in \cref{eq:Phi1_decomposed} yields, at the extraction radius,
\begin{equation}
    \phi_1^{00} = \frac{\sqrt{\pi} Q}{R_\text{ex}^2} \,,
\end{equation}
allowing the extraction of the total charge $Q$ from the measured $\phi_1^{00}$ at the chosen extraction radius.

\section{Initial Data}
\label{sec:initial-data}

The initial data considered in this work corresponds to asymptotically flat, spherically symmetric \glspl{BH} with resonant scalar hair in the model~\eqref{eq:action}. This type of solutions has been previously constructed and analyzed in~\cite{Herdeiro2020,Hong2020,Hong2020a}. Here, the procedure for generating such solutions is briefly summarized and further technical details can be found in the cited references.

The scalar field and the electromagnetic 4-potential are assumed to take the form
\begin{equation}
    \phi = \Phi(r) e^{-i \omega t} \,, \qquad A = U(r) dt \,,
    \label{eq:matter_ansatz}
\end{equation}
where $\Phi(r)$ denotes the radial profile of the scalar field, $\omega$ is its (real) oscillation frequency, and $U(r)$ is the electric potential. Throughout this section, $r$ denotes the isotropic radial coordinate.

To describe a spherically symmetric spacetime, the following metric ansatz in isotropic coordinates is adopted:
\begin{equation}
    \label{eq:metric-ansatz}
    ds^2 = -e^{2\mathcal{F}_0} \frac{S_0^2}{S_1^2} dt^2 + e^{2\mathcal{F}_1} S_1^4 d\Sigma^2 \,,
\end{equation}
where $d\Sigma^2$ is the line element of three-dimensional flat space, and $\mathcal{F}_0(r)$ and $\mathcal{F}_1(r)$ are undetermined functions. The auxiliary functions $S_0$ and $S_1$ are defined as
\begin{equation}
    S_0 \equiv 1 - \frac{r_h}{r} \,, \qquad S_1 \equiv 1 + \frac{r_h}{r} \,,
\end{equation}
with $r_h$ representing the horizon radius of the \gls{BH}.

Regularity of the energy-momentum tensor at the horizon imposes the resonance condition
\begin{equation}
    \omega = q U(r_h) \,.
\end{equation}
Imposing that the electric potential vanishes at spatial infinity, the bound-state condition for the scalar field yields an upper bound on the frequency,
\begin{equation}
    \label{eq:omega-upper-bound}
    \omega \leq \mu \,.
\end{equation}
These conditions are derived under the assumption that the gauge is chosen such that the electric potential $U(r)$ vanishes at spatial infinity.

Substituting the ansatz \cref{eq:metric-ansatz,eq:matter_ansatz} into the Einstein field equations, together with the matter sector equations, results in a system of ordinary differential equations for the four functions $\mathcal{F}_0(r)$, $\mathcal{F}_1(r)$, $\Phi(r)$, and $U(r)$. For a given model, specified by the parameters $q$ and $\mu,\lambda$ (for the potential \cref{eq:sextic-potential}) or $\mu,f,B$ (for the potential \cref{eq:axionic-potential}), solutions are obtained by varying the ansatz parameters $\omega$ and $r_h$.

\begin{figure}[h!]
    \centering
    \includegraphics[width=\linewidth]{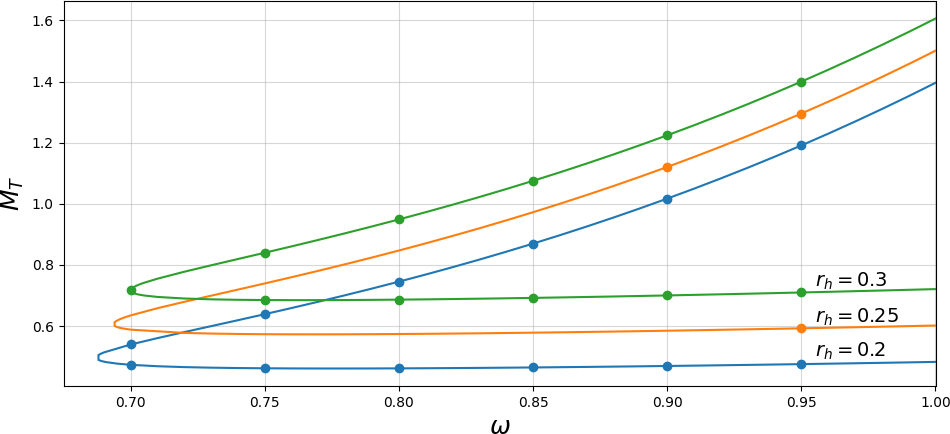}
    \caption{Solution curves for the Q-ball potential with $\mu = 1$, $\lambda = 2500$, and $q = 12$. $M_T$ is the total mass of the system. Dots indicate solutions that are evolved in the next section.}
    \label{fig:existence-curve}
\end{figure}

As in boson stars, a notable feature of the solution space is the existence of distinct branches for fixed values of $\omega$ and $r_h$. In this case we have an upper and a lower branch. The existence curves in the $(M_T, \omega)$ plane for the Q-ball potential \cref{eq:sextic-potential}, for several values of $r_h$, are displayed in \cref{fig:existence-curve}, where $M_T$ denotes the total mass of the system. A qualitatively similar plot is valid for the axionic potential \cref{eq:axionic-potential}.

\begin{figure}[h!]
    \centering
    \includegraphics[width=\linewidth]{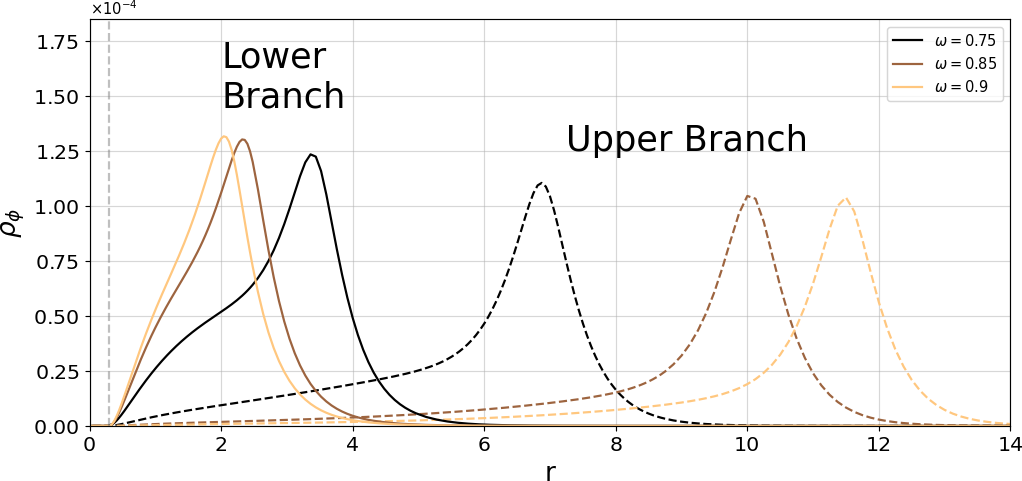}
    \caption{Radial profile of the scalar field density $\rho_\phi$ versus radius $r$ for solutions with $r_h = 0.2$ and three different values of $\omega$ for the axionic potential. For the lower branch (solid lines), increasing $\omega$ increases the compactness of the hair, while for the upper branch (dashed lines), the compactness decreases.}
    \label{fig:rho_phi-branches}
\end{figure}

Qualitatively, the upper branch solutions are characterized by less compact configurations, with the scalar hair extending further from the \gls{BH}. Along this branch, increasing $\omega$ leads to an increase in the spatial extent of the hair. In contrast, the lower branch exhibits the opposite behavior: increasing $\omega$ results in more compact configurations. This is illustrated in \cref{fig:rho_phi-branches}, which shows the density profile of the scalar field as a function of $r$ for representative values of $\omega$ belonging to both branches. The picture is qualitatively the same when considering the Q-ball potential.

For future reference, we note that the hair compactness does not vary significantly when fixing $\omega$ and varying $r_h$, along the lower branches.
\section{Results}
\label{sec:results}

\subsection{Numerical Setup}
\label{subsec:results-setup}

The evolutions are performed in units where the mass of the scalar field is set to $\mu = 1$. For the Q-ball potential we consider $\lambda = 2500$~\footnote{This large value of $\lambda$ allows the contributions of the quartic/sextic terms in the potential to be of the same order as that of the mass term.} and $q = 12$. For the axionic case, we set $B=0.22$, $f=0.01$ and $q=3$. We explored the parameter space region $0.2\leq r_h \leq 0.3$ and $\omega$ ranging from $0.7$ to $0.95$ in increments of $0.05$. The results presented in this section correspond to a representative subset of the simulations that were performed.

Time evolutions are performed using the Einstein Toolkit \cite{ETK,Zilhao:2013hia}, employing the \texttt{MagnetoScalar} thorn~\cite{Jaramillo2024}, the Carpet mesh refinement infrastructure~\cite{Carpet}, horizon tracking via \texttt{AHFinderDirect}~\cite{AHFinderDirect}, and metric evolution with BSSN~\cite{Baumgarte1998,Shibata1995} using the thorn \texttt{LeanBSSNMoL}~\cite{Canuda,Sperhake:2006cy}.

The simulations are performed on a $256 \times 256 \times 128$ grid with reflection symmetry imposed on the $xy$ plane. A box-in-a-box mesh refinement strategy with nine levels is employed, with a grid spacing of $h = 4$ on the coarsest mesh. The grid dynamically tracks the puncture location of the \gls{BH} during the evolution. Kreiss-Oliger dissipation is implemented using the ``continuous'' scheme described in \cite{Bozzola2021}. Spatial derivatives are computed using fourth-order finite differences, and time integration is carried out with a fourth-order Runge-Kutta method.

For the outer boundaries, we impose radiative boundary conditions using the \texttt{NewRad} infrastructure. Due to the usage of puncture coordinates for the time evolution, the horizon of the \gls{BH} does not require any special treatment.

Post-processing of the simulation data is performed with Kuibit \cite{Kuibit} and Visit \cite{Visit}. The analysis scripts and selected results are made publicly available in \cite{instability-qhairy}.

\begin{figure}[!t]
    \centering
    \subfigure[Initial configuration]{\includegraphics[width=0.45\columnwidth]{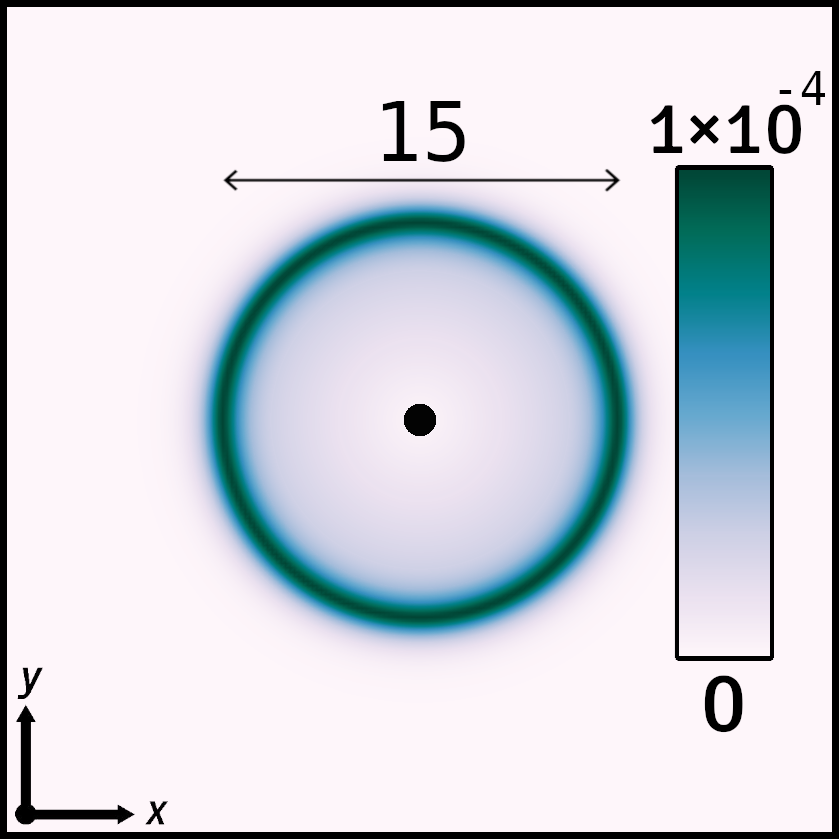}} \hfill
    \subfigure[Instability develops]{\includegraphics[width=0.45\columnwidth]{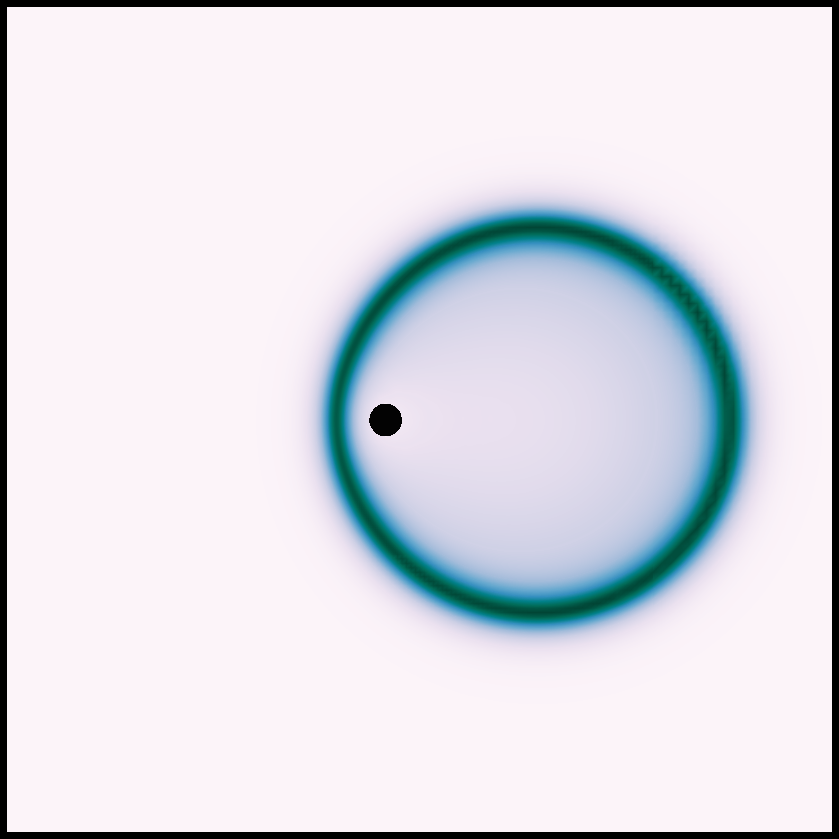}} \\
    \subfigure[\gls{BH} is expelled by the hair]{\includegraphics[width=0.45\columnwidth]{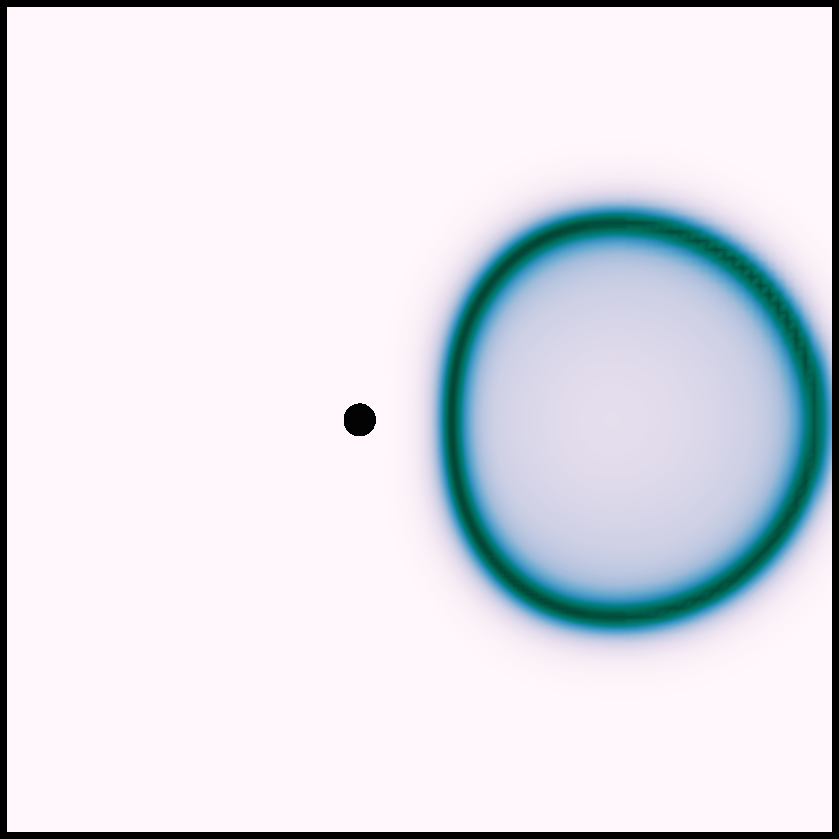}} \hfill
    \subfigure[Final state, \gls{BS} plus \gls{BH}]{\includegraphics[width=0.45\columnwidth]{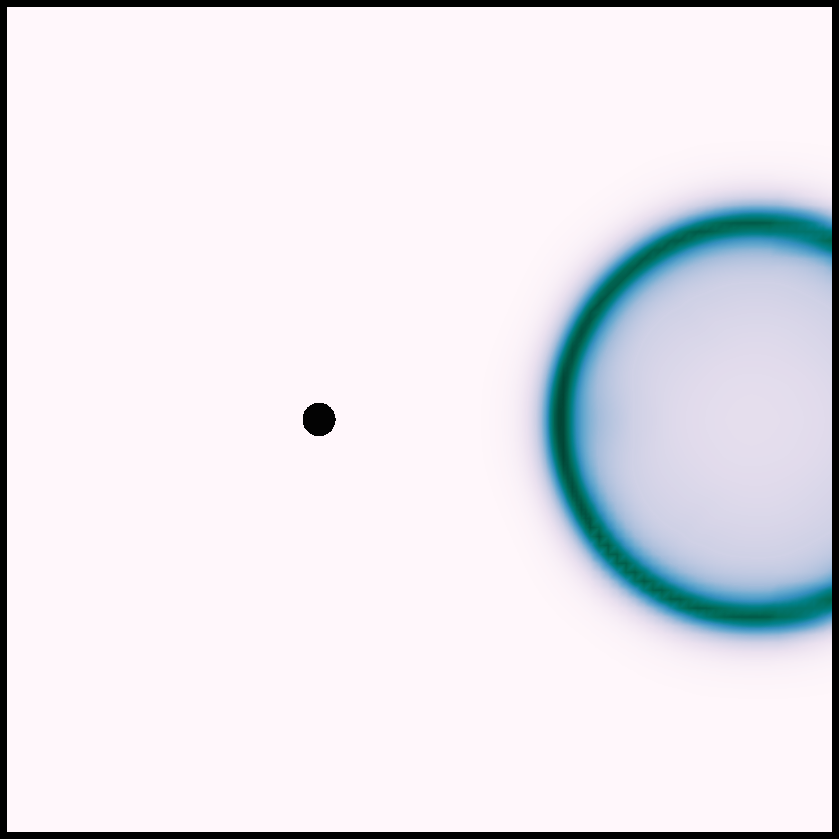}}
    \caption{Snapshots of the normal density of the scalar field with axionic potential in the $xy$ plane for \textit{fission}, with $r_h = 0.3$ and $\omega = 0.75$ in the upper branch. The final state is a stable \gls{BS} and a hairless \gls{BH} with opposite momenta.}
    \label{fig:fission-snapshots}
\end{figure}

\begin{figure}[!t]
    \centering
    \subfigure[Initial configuration]{\includegraphics[width=0.45\columnwidth]{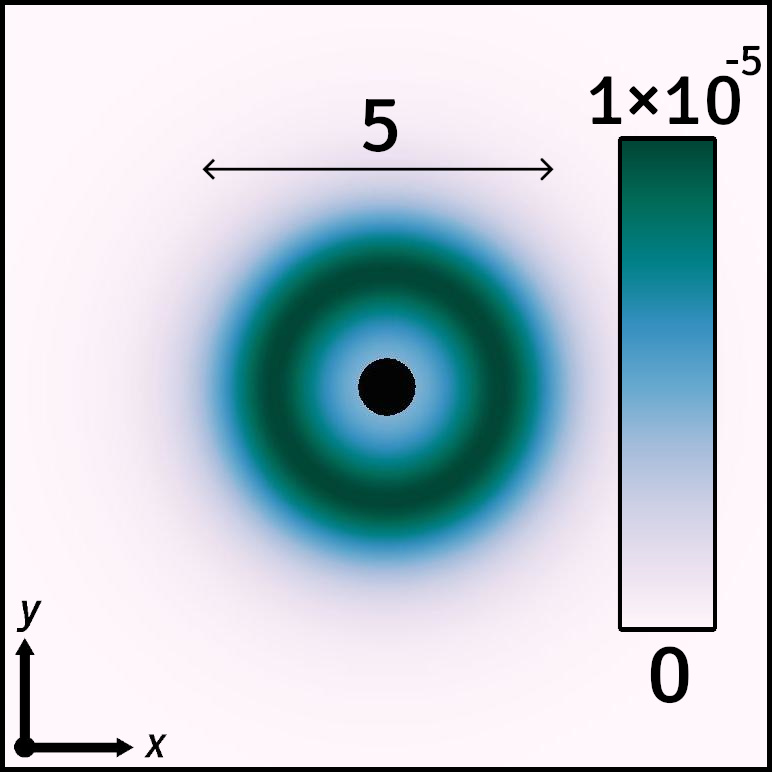}} \hfill
    \subfigure[Instability develops]{\includegraphics[width=0.45\columnwidth]{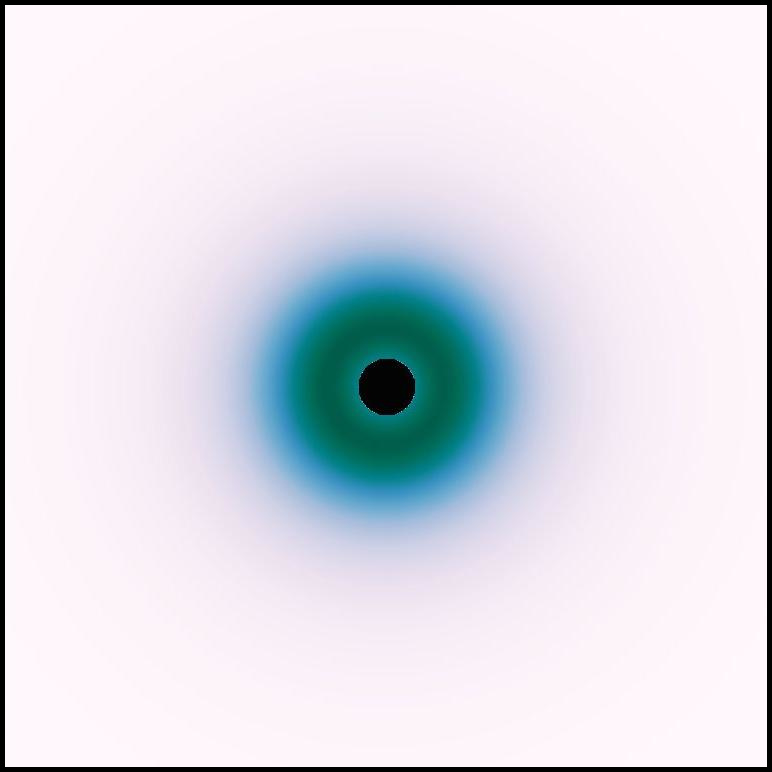}} \\
    \subfigure[Hair is getting absorbed]{\includegraphics[width=0.45\columnwidth]{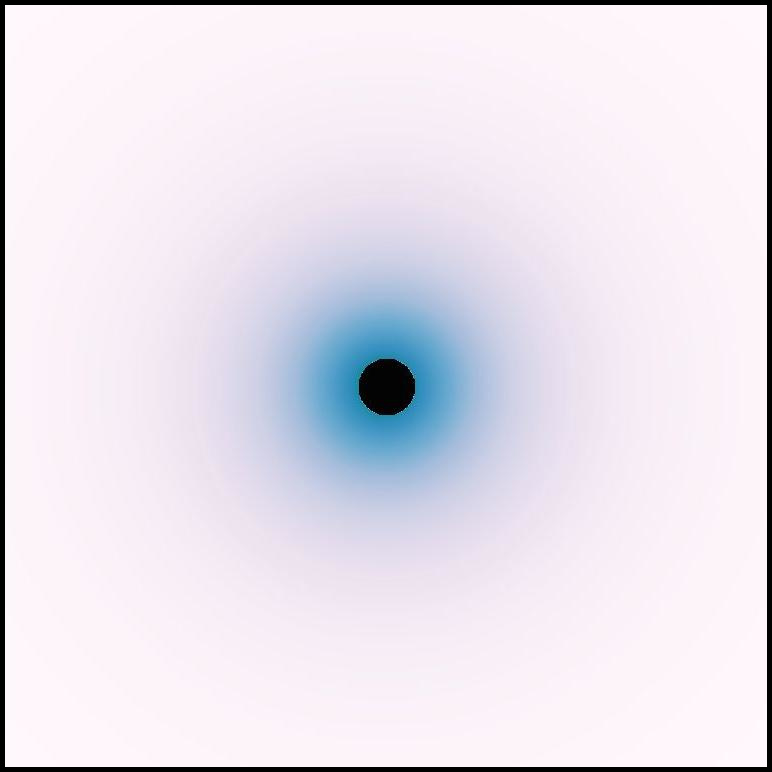}} \hfill
    \subfigure[Final state, a hairless \gls{BH}]{\includegraphics[width=0.45\columnwidth]{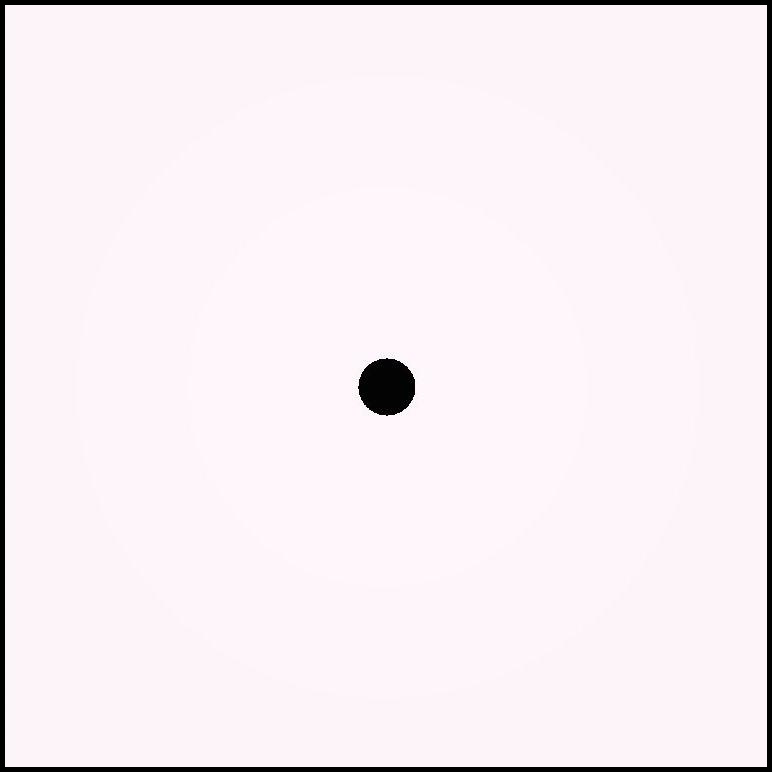}}
    \caption{Snapshots of the normal density of the scalar field with Q-Ball potential in the $xy$ plane for \textit{absorption}, with $r_h = 0.2$ and $\omega = 0.95$ in the lower branch. The final state is characterized by a hairless \gls{BH}.}
    \label{fig:absorption-snapshots}
\end{figure}

Details regarding the convergence of the numerical setup are provided in \cref{app:convergence}.

\subsection{Time Evolutions}
\label{subsec:results-evolution}

Evolving these solutions reveals two qualitatively distinct outcomes: (i) \textit{absorption}, in which the hair is absorbed by the \gls{BH}, and (ii) \textit{fission}, already discussed in~\cite{Nicoules2025}, where the \gls{BH} is expelled from the hair resulting in a stable (charged) \gls{BS} and a hairless (charged) \gls{BH}. Representative snapshots of these decay mechanisms are displayed in \cref{fig:fission-snapshots} for fission and \cref{fig:absorption-snapshots} for absorption.

The fate of each solution seems to be correlated with the compactness of the hair. Solutions with more compact hair, typically found in the lower branch at higher $\omega$, undergo absorption. In contrast, less compact configurations, located in the upper branch or in the lower branch with smaller $\omega$, exhibit the fission instability. We emphasize that these behaviors are universal for both potentials considered.

\begin{figure}[thpb]
    \centering
    \subfigure{\includegraphics[width=0.49\columnwidth]{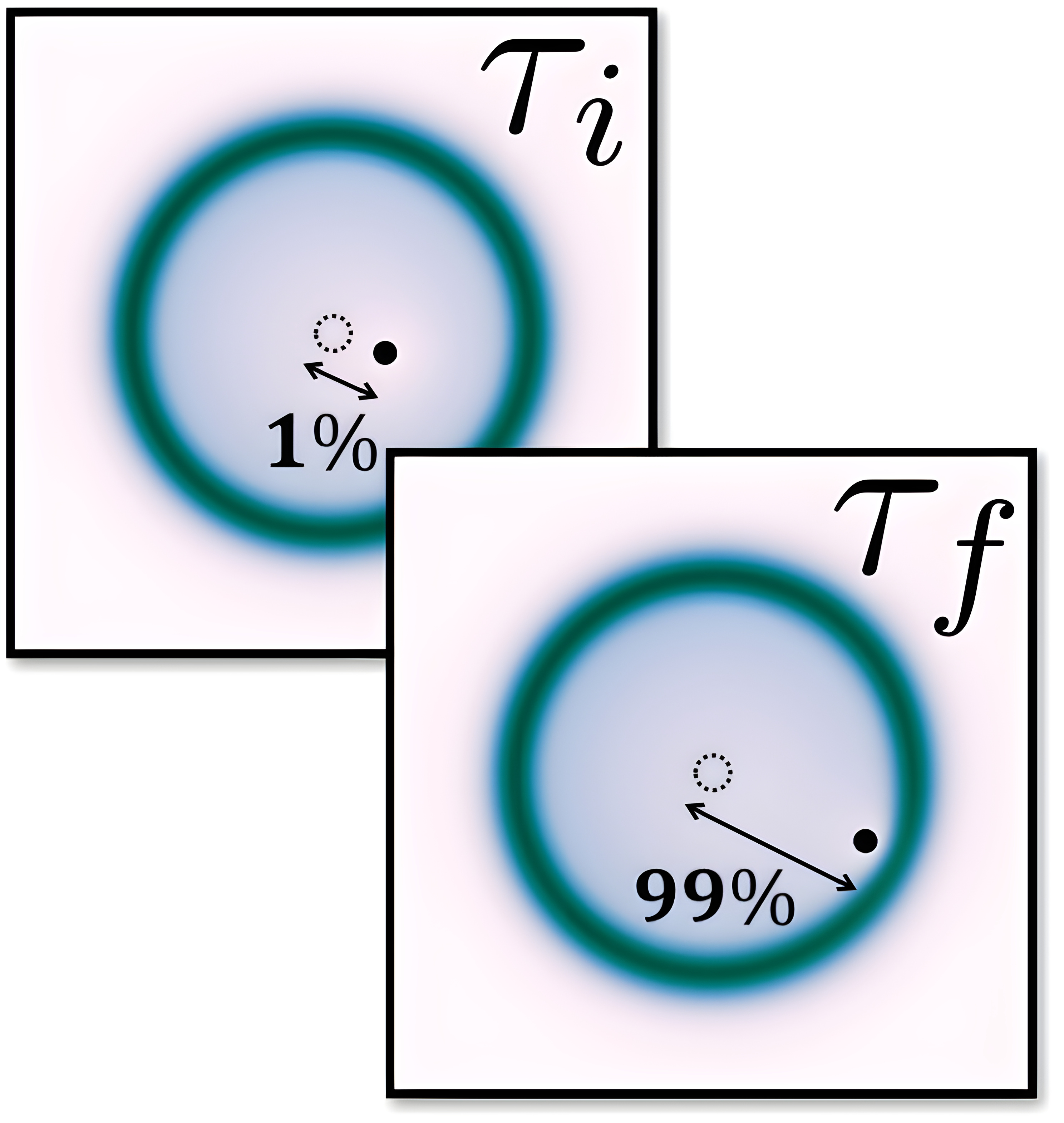}} \hfill
    \subfigure{\includegraphics[width=0.49\columnwidth]{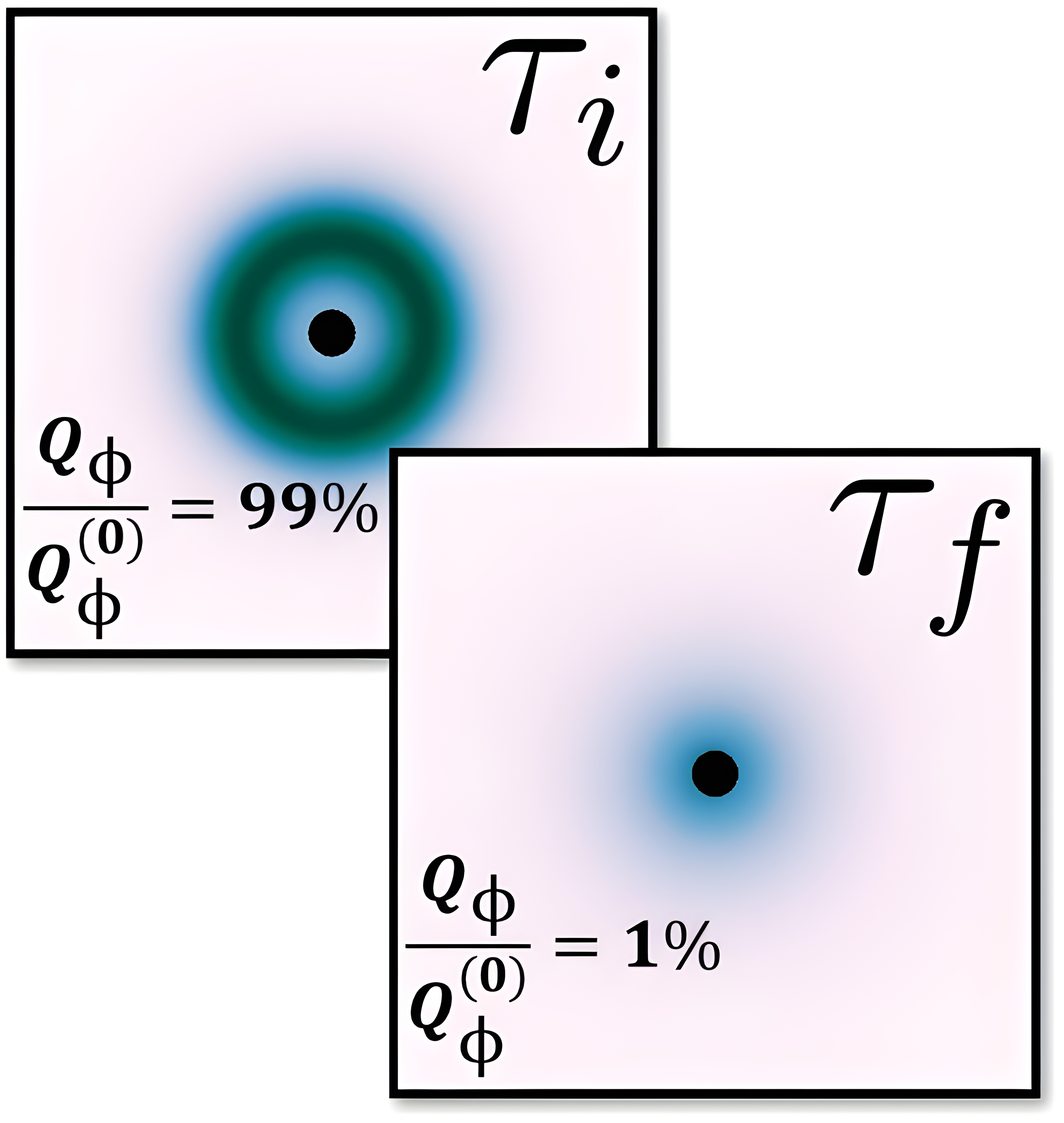}} \\
    \caption{Schematic representation of the definitions for $\tau_i$ and $\tau_f$ for \textit{fission} (left) and for \textit{absorption} (right).}
    \label{fig:def-timescales}
\end{figure}

To quantitatively characterize the lifetime of each solution, we define two instants of time, $\tau_i$ and $\tau_f$, represented schematically for each decay mechanism in \cref{fig:def-timescales}. Additionally, we define the quantity $\Delta \tau \equiv \tau_f - \tau_i$, that we interpret as the timescale of the process.

For \textit{fission}, we define a sphere of radius $R_i$ as the sphere containing 1\% of the total charge of the scalar field, and $R_f$ as the radius of a sphere containing 99\% of the total charge of the scalar field\footnote{The radius of the spherical surface is evaluated at $t=0$. Since the scalar hair remains approximately spherically symmetric until the centroid is expelled, this radius is treated as constant.}%
.
We define $\tau_i$ as the moment in time when the centroid of the \gls{BH} crosses $R_i$, and $\tau_f$ as the time it crosses $R_f$.

As for \textit{absorption}, $\tau_i$ is defined as the time when the charge of the scalar field $Q_\phi$ drops to $99\%$ of its initial value, and $\tau_f$ as the time when $Q_\phi$ reaches $1\%$ of its initial value.

We verified that using different values in the definition of $\tau_i$ and $\tau_f$ for either scenarios (e.g. $5\%$ and $95\%$ or $10\%$ and $90\%$), does not qualitatively affect the conclusions. For more information, refer to \cref{app:charge_threshold}.

Since we do not add any perturbation to our system, the only source of the instability is numerical noise. This means that, for the same physical system, but for different numerical setups (e.g. resolution, dissipation, etc.), the values of both $\tau_i$ and $\tau_f$ vary. However, the value of $\Delta \tau$ remains consistent across different numerical configurations. Because of this, our analysis will only focus on $\Delta \tau$, as it is the only physically meaningful timescale in this context.

\begin{figure}[t!]
    \centering
    \includegraphics[width=\linewidth]{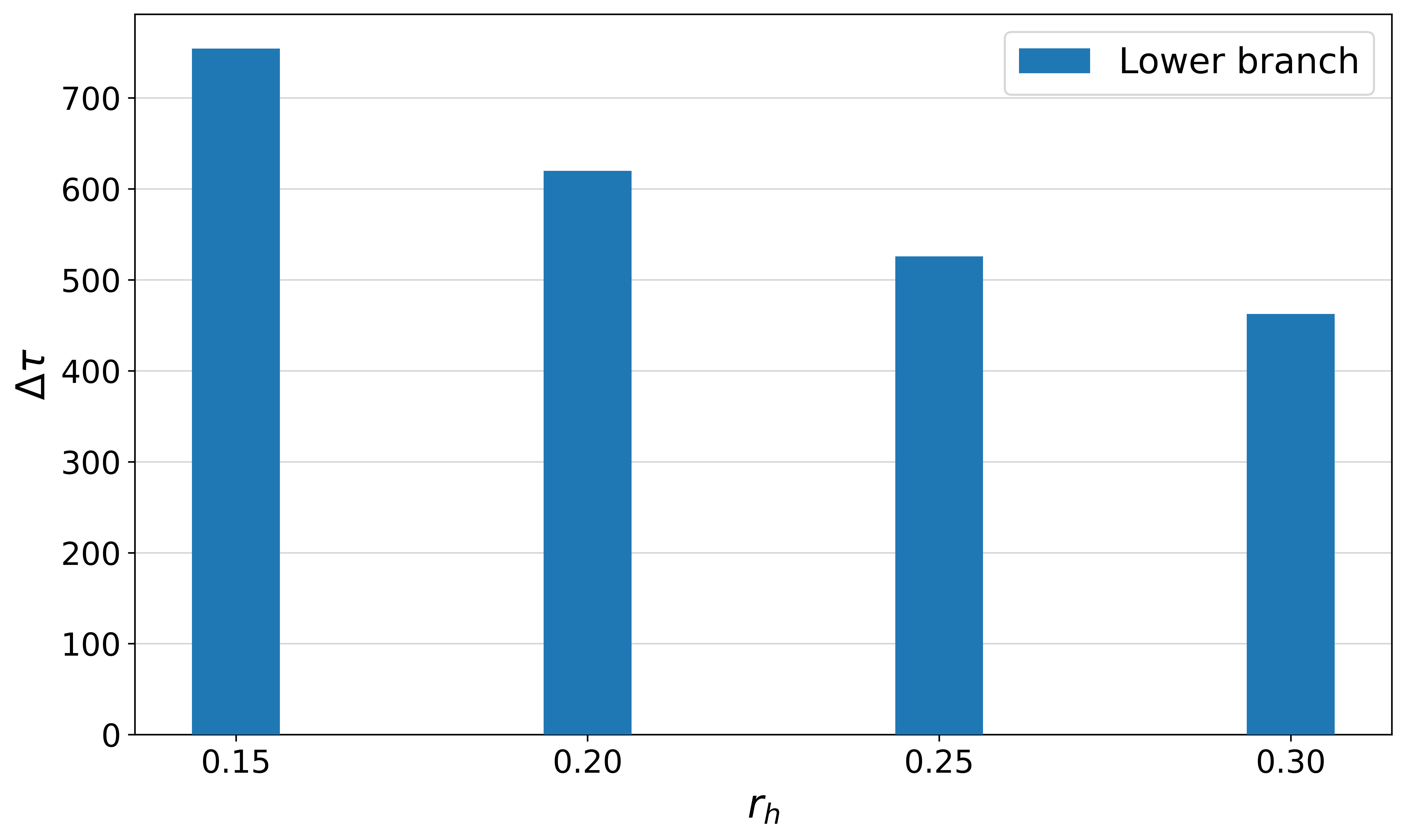}
    \caption{The value of $\Delta \tau$ versus $r_h$ for \textit{absorption}, considering the Q-ball potential and $\omega = 0.95$. All solutions exhibiting this instability are in the lower branch.}
    \label{fig:timescales-qball-absorption}
\end{figure}

\begin{figure}[t!]
    \centering
    \includegraphics[width=\linewidth]{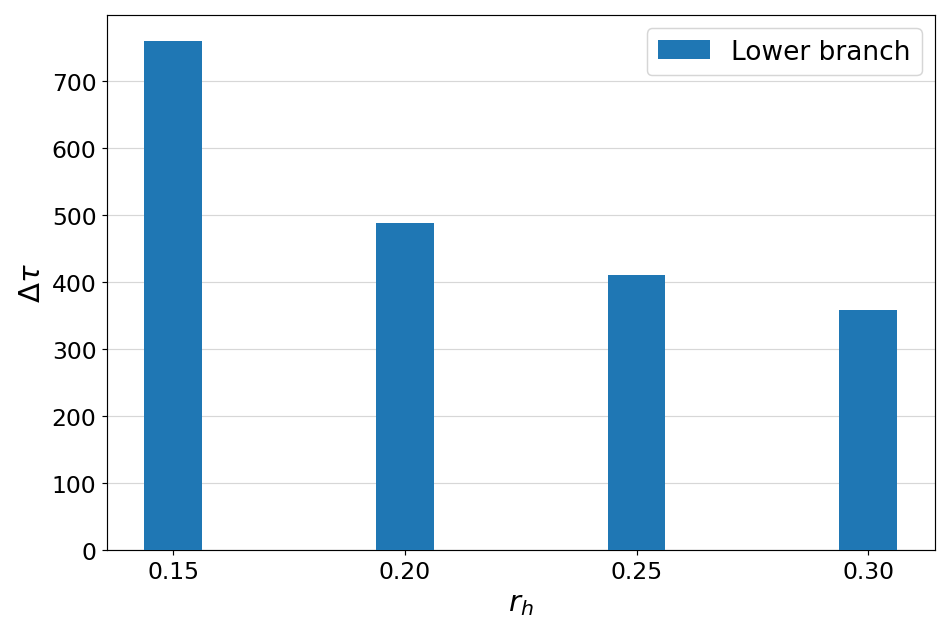}
    \caption{The value of $\Delta \tau$ versus $r_h$ for \textit{absorption}, considering the axionic potential and $\omega = 0.95$. All solutions exhibiting this instability are in the lower branch.}
    \label{fig:timescales-axionic-absorption}
\end{figure}

In \cref{fig:timescales-qball-absorption}, we present the timescale $\Delta \tau$ for the absorption process as a function of $r_h$, considering Q-ball potential and $\omega = 0.95$. Here, $r_h$ is chosen as the reference parameter since the absorption instability is observed exclusively for solutions with more compact hair, typically found in the lower branch. For these cases, $r_h$ is varied between solutions in the lower branch. The results show that $\Delta \tau$ decreases monotonically with increasing $r_h$, indicating that (compact hair) configurations with larger \glspl{BH} are more unstable (at fixed $\omega$). For the axionic potential, presented in \cref{fig:timescales-axionic-absorption}, we observe the same behavior: increasing $r_h$ decreases the value of $\Delta \tau$ monotonically.

\begin{figure}[h!]
    \centering
    \includegraphics[width=\linewidth]{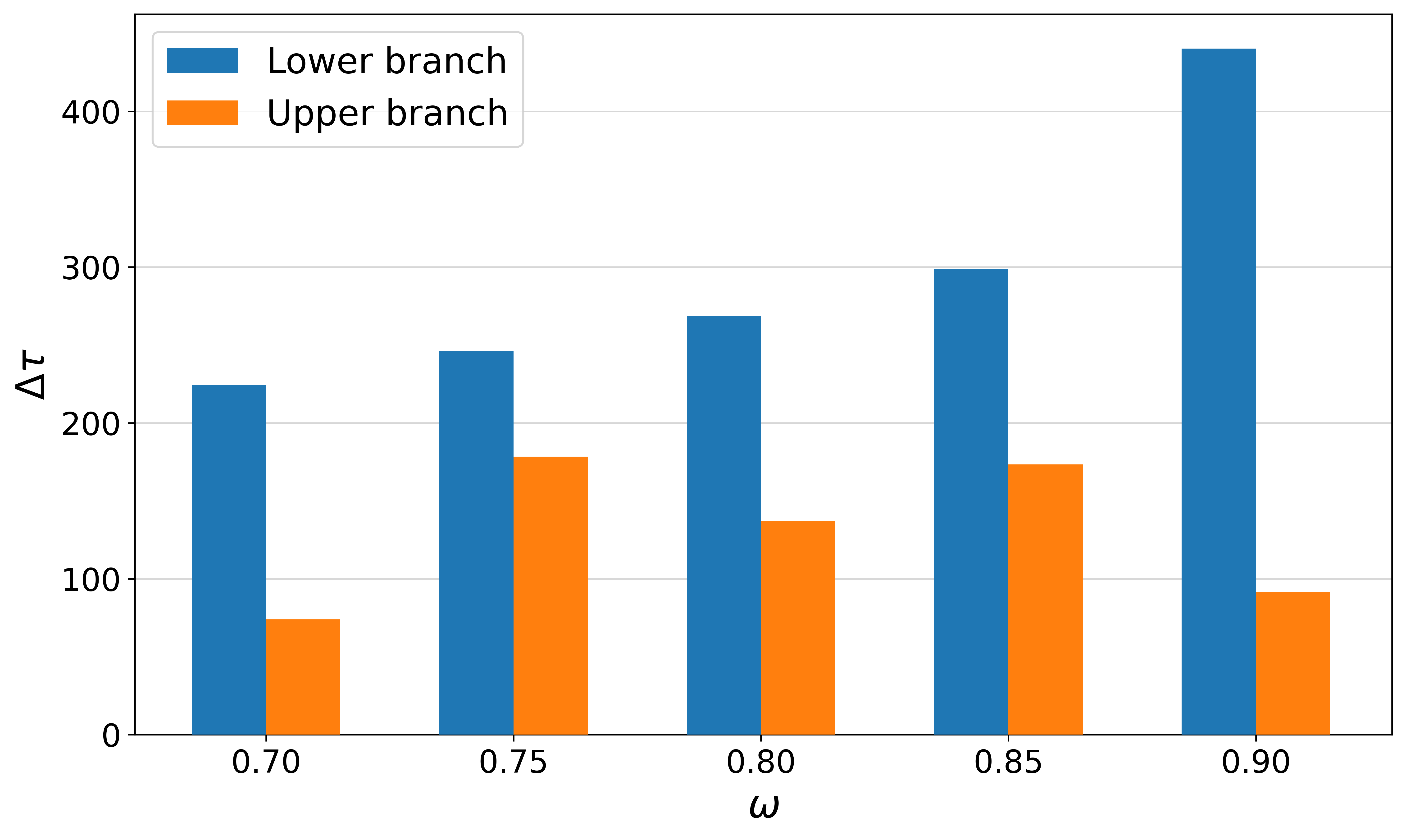}
    \caption{The value of $\Delta \tau$ versus $\omega$ for \textit{fission}, considering the Q-ball potential and $r_h = 0.2$.}
    \label{fig:timescales-qball-fission}
\end{figure}

\begin{figure}[h!]
    \centering
    \includegraphics[width=\linewidth]{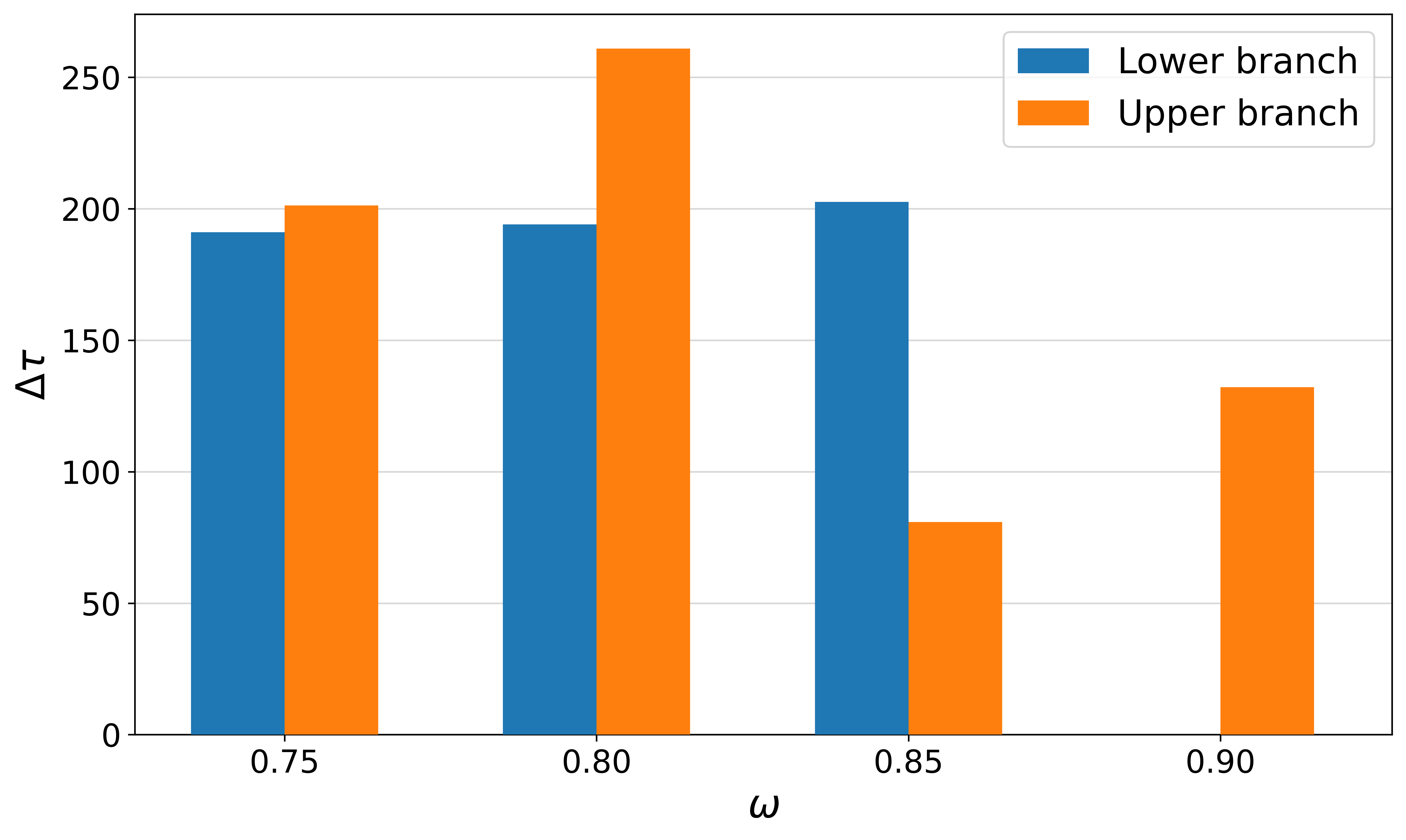}
    \caption{The value of $\Delta \tau$ versus $\omega$ for \textit{fission}, considering the axionic potential and $r_h = 0.3$.}
    \label{fig:timescales-axionic-fission}
\end{figure}

In \cref{fig:timescales-qball-fission} we present $\Delta \tau$ for the fission instability versus $\omega$, assuming a Q-ball potential and fixing $r_h = 0.2$. We can see that $\Delta \tau$ increases with $\omega$ in the lower branch, showing that more compact solutions take longer to decay than less compact solutions. In the upper branch, $\Delta \tau$ exhibits a non-monotonic dependence on $\omega$, reflecting the influence of additional parameters such as mass and charge for the stability of the configuration. For the axionic potential, \cref{fig:timescales-axionic-fission} demonstrates that the instability timescale $\Delta \tau$ exhibits minimal dependence on the frequency $\omega$ along the lower branch, whereas the upper branch displays non-monotonic behavior consistent with the results obtained considering the Q-ball potential.

In both potentials, we see a similar pattern: more compact solutions exhibit the absorption instability, while less compact solutions exhibit the fission instability. Notably, the values of $\Delta \tau$ are of the same order between both potentials, indicating that the timescale of the instability is independent of the specific form of the scalar potential.

\section{Final Remarks}
\label{sec:final-remarks}

In this work, we studied the stability of \glspl{BH} with resonant hair using fully nonlinear $3+1$ numerical relativity simulations. Although previous work in spherical symmetry indicated stability, the present analysis has identified two distinct decay mechanisms arising from non-spherically symmetric instabilities: one in which the \gls{BH} absorbs the surrounding hair, and another in which the \gls{BH} is expelled from the scalar hair, leaving behind a stable \gls{BS}. In both cases, the loss of spherical symmetry seems to be a key ingredient for the instability to develop.

It was observed that solutions with more compact hair tend to absorb the scalar field, whereas those with less compact configurations are more susceptible to the ejection of the \gls{BH}. In some regions of parameter space, we found that the timescale of the decay was correlated with the compactness of the scalar hair. However, this correlation was not universal, and competing factors such as the mass and the charge of the hair and those of the \gls{BH} also play a role in determining the timescale of the instability. A systematic exploration of the parameter space is required to better understand the dependence of the timescales of the instability.

These results indicate that the observed instabilities are generic properties of \glspl{BH} with resonant hair\footnote{A process with some similarities has been observed in AdS for spherically symmetric black holes in \cite{Zhuan2024}.}. This instability is caused by the unstable balance between gravitational pull and electromagnetic repulsion between the hair and the \gls{BH}, which we expect to affect a wider region of parameter space and to be largely independent of the specific scalar potential.

\begin{acknowledgments}

This work is supported by the Center for Research and Development in Mathematics and Applications (CIDMA) (\url{https://ror.org/05pm2mw36}) under the Portuguese Foundation for Science and Technology 
(FCT -- Fundaç\~ao para a Ci\^encia e a Tecnologia, \url{https://ror.org/00snfqn58}), Grants UID/04106/2025 (\url{https://doi.org/10.54499/UID/04106/2025}) and UID/PRR/04106/2025 (\url{https://doi.org/10.54499/UID/PRR/04106/2025}), as well as the projects: Horizon Europe staff exchange (SE) programme HORIZON-MSCA2021-SE-01 Grant No.\ NewFunFiCO-101086251 and 2022.04560.PTDC (\url{https://doi.org/10.54499/2022.04560.PTDC}).
J.F.\ is funded by FCT through project 2023.04333.BD (\url{https://doi.org/10.54499/2023.04333.BD}).
The authors thankfully acknowledge computational resources from RES provided by BSC (MareNostrum) through projects FI-2024-2-0012, FI-2024-3-0007, POR021PROD, and by IFCA (Altamira) through project FI-2025-1-0011.
Computational resources were also provided via FCT through projects 2025.09498.CPCA.A3, 2024.07872.CPCA.A2 (DOI: 10.54499/2024.07872.CPCA.A2 \url{https://doi.org/10.54499/2024.07872.CPCA.A2}) at Deucalion supercomputer, jointly funded by EuroHPC JU and Portugal,
and at MareNostrum through project 2024.07059.CPCA.A3 (DOI: 10.54499/2024.07059.CPCA.A3 \url{https://doi.org/10.54499/2024.07059.CPCA.A3}).

\end{acknowledgments}

\appendix
\section{Dependence of Timescales on Scalar Field Charge Threshold}
\label{app:charge_threshold}

\begin{figure}[h!]
    \centering
    \includegraphics[width=0.97\linewidth]{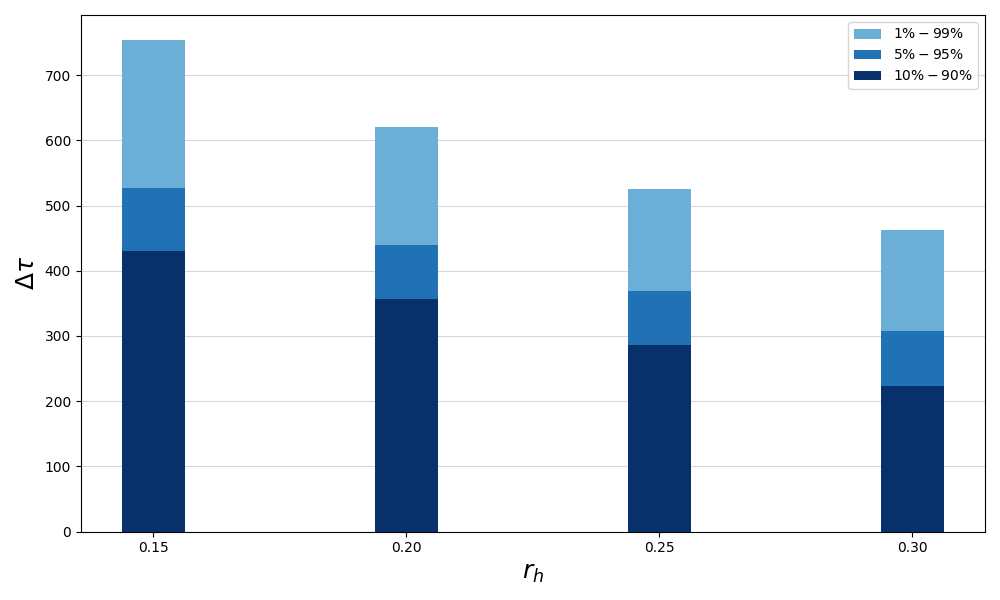}
    \caption{Value of $\Delta \tau$ versus $r_h$ for the \textit{absorption} scenario in Q-ball potential with $\omega = 0.95$, for different thresholds. All solutions exhibiting this instability are in the lower branch.}
    \label{fig:threshold-qball-absorption}
\end{figure}

\begin{figure}[h!]
    \centering
    \includegraphics[width=0.97\linewidth]{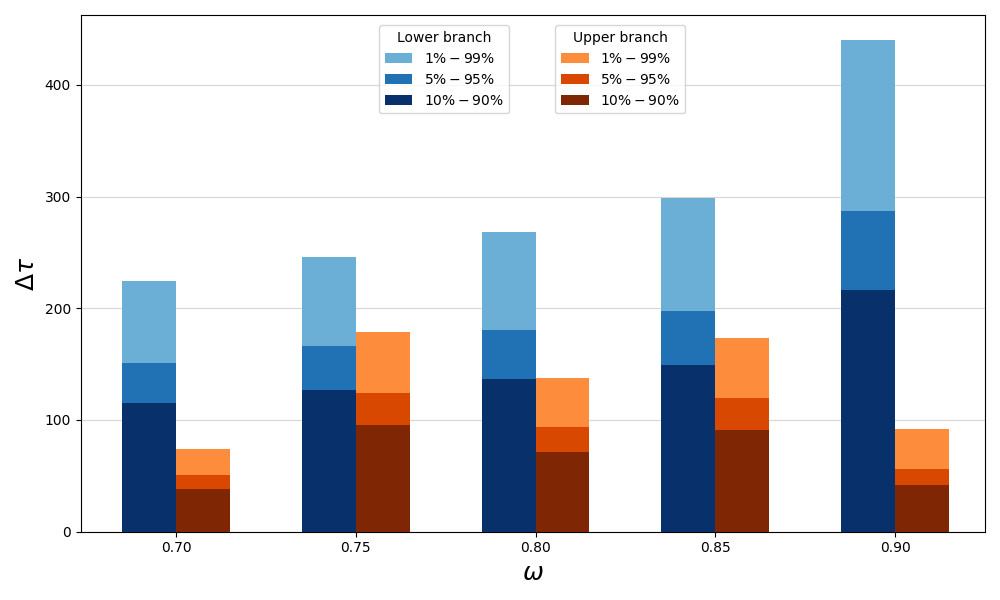}
    \caption{Value of $\Delta \tau$ versus $\omega$ for the \textit{fission} scenario in Q-ball potential with $r_h = 0.2$, for different thresholds.}
    \label{fig:threshold-qball-fission}
\end{figure}

We here investigate how the choice of thresholds for defining the timescales $\tau_i$ and $\tau_f$ affects the value of $\Delta \tau$. We consider three different sets of thresholds: \textit{(i)} $1\%$ and $99\%$, \textit{(ii)} $5\%$ and $95\%$, and \textit{(iii)} $10\%$ and $90\%$. The results for the Q-ball potential are summarized in \cref{fig:threshold-qball-absorption}, for the absorption scenario, and in \cref{fig:threshold-qball-fission} for the fission scenario.

\begin{figure}[h!]
    \centering
    \includegraphics[width=0.97\linewidth]{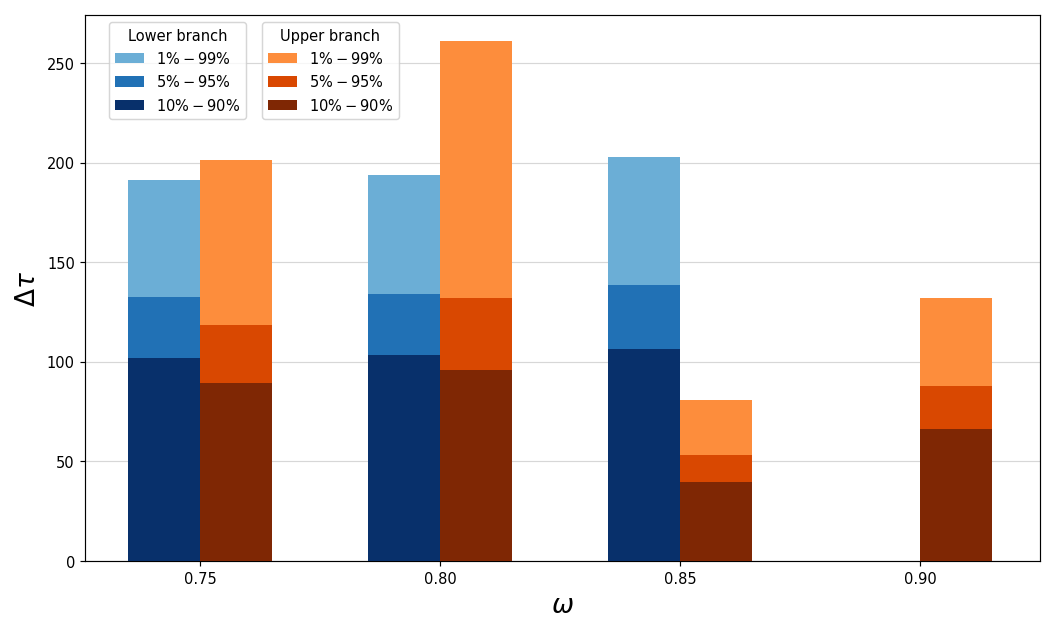}
    \caption{Value of $\Delta \tau$ versus $\omega$ for the \textit{fission} scenario in axionic potential with $r_h = 0.3$, for different thresholds.}
    \label{fig:threshold-axionic-fission}
\end{figure}

\begin{figure}[h!]
    \centering
    \includegraphics[width=0.97\linewidth]{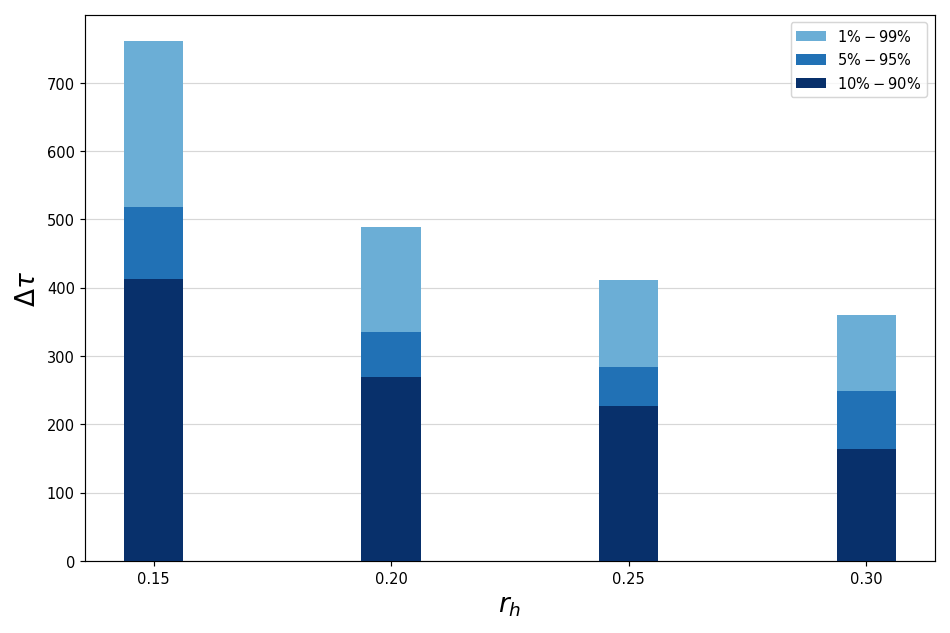}
    \caption{Value of $\Delta \tau$ versus $r_h$ for the \textit{absorption} scenario in axionic potential with $\omega = 0.95$, for different thresholds.}
    \label{fig:threshold-axionic-absorption}
\end{figure}

For the axionic potential, the results are presented in \cref{fig:threshold-axionic-fission} for the fission scenario and in \cref{fig:threshold-axionic-absorption} for the absorption scenario.

As illustrated in the figures, the choice of thresholds does not change the qualitative behavior of $\Delta \tau$, as the trends observed in the main text remain consistent across different threshold values.

\section{Convergence}
\label{app:convergence}

Here we present some convergence results that validate the numerical evolution. We monitor the Hamiltonian constraint
\begin{equation}
    \mathcal{H} \equiv \mathcal{R} + K^2 - K_{i j} K^{i j} - 16 \pi \rho \,,
\end{equation}
where $\mathcal{R}$ is the Ricci scalar of the spatial metric, $\rho \equiv n_a n_b T^{a b}$ is the normal energy density and $K$ the trace of the extrinsic curvature $K_{i j}$ that is defined as
\begin{equation}
    K_{ij} \equiv - D_i n_j \,.
\end{equation}

\begin{figure}[]
    \centering
    \includegraphics[width=\linewidth]{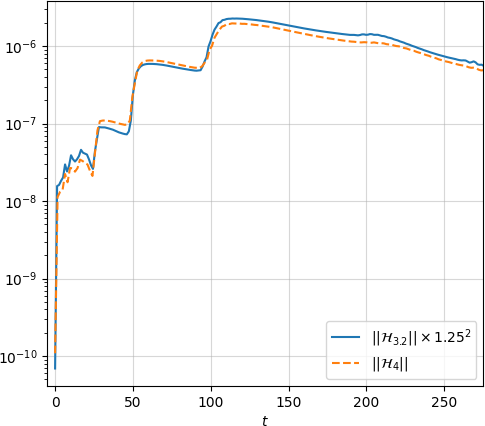}
    \caption{Convergence to zero of the norm-2 of the Hamiltonian constraint $\mathcal{H}$ for Q-ball potential with $r_h = 0.3$ and $\omega = 0.95$, in the lower branch, corresponding to the absorption process.}
    \label{fig:hamiltonian-absorption}
\end{figure}

For the case of \textit{absorption}, we run the same simulation with two different resolutions and perform a convergence analysis of the norm-2 of the Hamiltonian. We show that the norm-2 of the Hamiltonian constraint $\mathcal{H}$ exhibits second-order convergence to zero. This behavior is illustrated in \cref{fig:hamiltonian-absorption}.

\begin{figure*}[]
    \centering
    \subfigure[Hamiltonian constraint $\mathcal{H}$ before fission takes place.]
    {\includegraphics[width=0.45\textwidth]{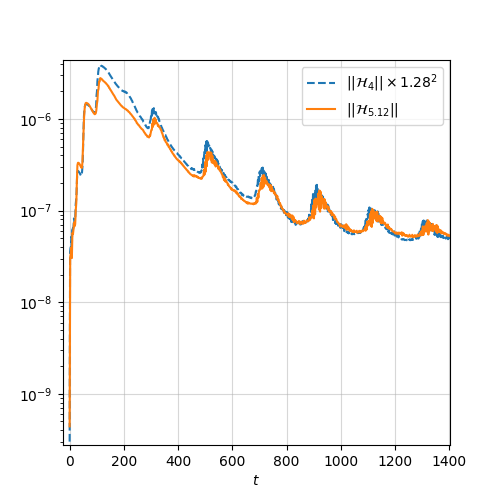}}
    \subfigure[Hamiltonian constraint $\mathcal{H}$ during the fission process.]
    {\includegraphics[width=0.45\textwidth]{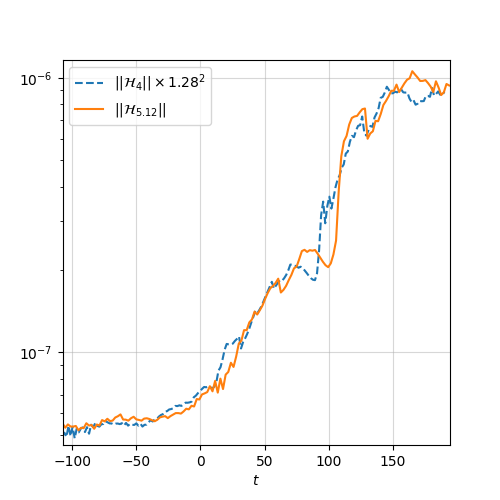}}
    \caption{Convergence to zero of the norm-2 of the Hamiltonian constraint $\mathcal{H}$ for Q-ball potential with $r_h = 0.3$ and $\omega = 0.7$, in the upper branch. The plot on the right is time shifted to match the position of the punctures, such that $t=0$ corresponds to the time the puncture leaves the region in which most of the hair is located.}
    \label{fig:hamiltonian-fission}
\end{figure*}

For \textit{fission} we also observe second order convergence to zero of the norm-2 of the Hamiltonian constraint (left plot of \cref{fig:hamiltonian-fission}). Since instabilities are sourced by numerical noise, the time at which the \gls{BH} starts moving, $\tau_i$, differs between different resolutions. If we re-scale the time in this regime appropriately, we can see that the Hamiltonian constraint still converges with second order (right plot of \cref{fig:hamiltonian-fission}).

The convergence observed is second-order, consistent with the second-order prolongation operations used by Carpet. Fourth-order convergence is obtained at $t = 0$, however, as expected.

\bibliography{bibliography}

\end{document}